\documentclass{article}
\usepackage{amsmath}
\usepackage{amsthm}
\usepackage{amsfonts}
\usepackage{amssymb}
\usepackage{graphicx}
\usepackage{cmap}
\usepackage[T1]{fontenc}
\usepackage[utf8]{inputenc}
\usepackage[english]{babel}

\newcommand{\beq}{\begin{equation}}
\newcommand{\eeq}{\end{equation}}

\begin{document}
\title{On the origin of self-oscillations in large systems.}

\author{Daniele De Martino \\
\\
\\
Institute of Science and Technology Austria,\\ Am Campus 1, A-3400 Klosterneuburg, Austria}
\maketitle
\abstract{
In this article is shown that large systems endowing phase coexistence display self-oscillations in presence of linear feedback between the control and order parameters, where an Andronov-Hopf bifurcation takes over the phase transition. This is simply illustrated through the mean field Landau theory whose feedback dynamics turns out to be described  by the Van der Pol equation and it is then validated for the fully connected Ising model following heat bath dynamics. Despite its simplicity, this theory accounts potentially for a rich range of phenomena: here it is  applied to describe in a stylized way  i) excess demand-price cycles due to strong herding  in  a simple agent-based market model; ii) congestion waves in  queuing networks  triggered by users feedback to delays in overloaded conditions; iii)  metabolic network oscillations resulting from cell growth control in a bistable phenotypic landscape.       
}

\section*{Introduction}
From  clocks to motors \cite{andronov1966theory}, many  musical instruments \cite{rayleigh1896theory, de2007analogical} (including our throat \cite{buccheri2016experimental}), pumping hearts \cite{van1928lxxii} \& firing neurons \cite{izhikevich2007dynamical}, many oscillatory systems are self-sustained \cite{andronov1966theory}. 
Despite their many applications \cite{guerra2009coupled} to describe natural and engineering systems, a thorough physical understanding of self-oscillations is lacking \cite{jenkins2013self}, in particular their emergence as a collective behavior in large systems, like the ones studied by statistical mechanics.  This in turn hampers  thermodynamical analysis, an aspect that started to puzzle scholars since the first seminal articles on the subject, where self oscillations were provokingly regarded as a form of perpetual motion \cite{airy1830certain}.  
Oscillatory collective behaviors of many interacting units have been mainly investigated from the point of view of their  synchronization where the interacting units are already assumed as linear oscillators \cite{kuramoto1975self, winfree1967biological} or self-oscillators \cite{de2009limit, di2017ginzburg}. 
In this work a different route is taken and a general criterion will be given for the onset of self-oscillations in large systems in a fully self-organized way, without postulating that the elementary units are oscillators. Rigorous criteria exist for the development of self  oscillations in dynamical systems \cite{guckenheimer2013nonlinear}. From a more physical viewpoint, in particular in the context of electrical engineering, self oscillations are triggered for a workload corresponding to the ``negative resistence'' part of the  current-voltage characteristic curve of  active devices \cite{andronov1966theory}. 
The key idea of this work is that systems with many interacting degrees of freedom that show phase coexistence, upon treating the equation of states  on equal footing of characteristic curves, they develop self-oscillations in presence of feedback between the control and order parameters that try to force them  on thermodynamically unstable branches. This is illustrated in a {\em Gedankenexperiment} in the next section where the feedback dynamics   of the Landau mean field theory turns out to be described by the Van der Pol oscillator, a prediction that is successfully tested for the fully connected Ising model subject to heat bath dynamics.
While in the Ising model the feedback is artifically introduced for illustrative purposes,  in the following sections several different phenomena are considered in a stylized way in which the feedback is fully dynamically justified, namely  i) excess demand-price cycles due to strong herding  in  a simple agent-based market model; ii) congestion waves in  queuing networks  triggered by users feedback to delays in overloaded conditions; iii)  metabolic network oscillations resulting from cell growth control in a bistable phenotypic landscape. In the conclusions results are summarized and several interesting outlooks that stem from this work are pointed out.

\section*{Self-oscillating Ising model}
Consider the {\em gedankenexperiment} depicted in figure 1. 
We have a  magnet in equilibrium in a given external magnetic field $h$ and thermal bath of temperature $T<T_c$  (where $T_c$ is the critical temperature for the para-ferromagnetic transition) and 
we measure its magnetization $m(T,h)$: can  we control and set it to $m=0$ by tuning the external magnetic field with a simple linear negative feedback $dh \propto -m dt$, at fixed temperature?       
\begin{figure}[h!!!!!]\label{fig1}
\begin{center}
\includegraphics*[width=1\textwidth,angle=0]{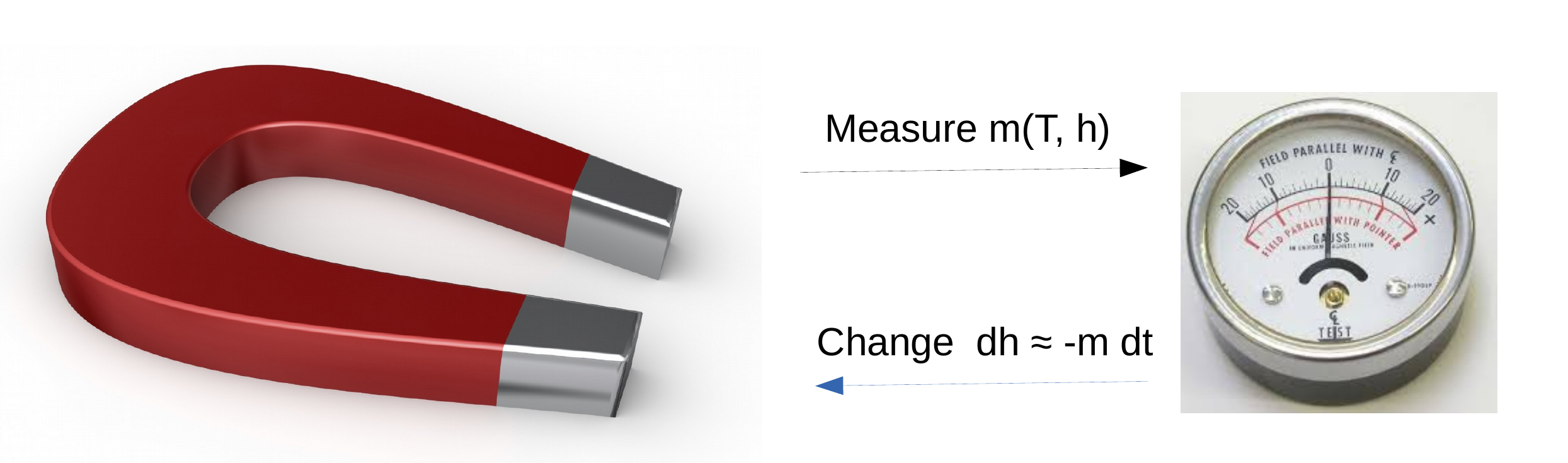}
\caption{{\em Gedankenexperiment}: can we control and set $m=0$ in magnet below the critical temperature $T<T_c$ using the external field $h$ with a simple linear negative feedback?}
\end{center}
\end{figure}
In the following it will be shown that this is not the case at least for the mean field case.
In order to address the question formally, we
consider the Landau expansion of the free energy density in the magnetization $m$ \cite{landau1969statistical} with parameters  $\beta=1/T$ and  $h$:
\begin{equation}
f(m) \sim -\frac{1}{2} (\beta -1) m^2 - m h +\frac{1}{12} m^4 + \dots
\end{equation}
Equilibrium relaxation dynamics follows approximately the phenomenological equation (linear response, where time unit coincides with the typical one) \cite{zwanzig2001nonequilibrium}  
\begin{equation}
\dot{m} = -\frac{\partial f}{\partial m}.
\end{equation}
The extrema of the free energy  are thus equilibrium states of the system: without external magnetic field $h=0$ we have $m=0$ as the unique stable equilibrium point (node) as soon as $\beta<1$, whereas for $\beta>1$ this becomes unstable while other two nodes appears at  
$m=\pm\sqrt{3(\beta-1)}$ that are stable equilibrium points (spontaneous symmetry breaking). 
Let us now add the linear negative feedback between the external field and the magnetization with timescale $\tau$, i.e. $\tau \dot{h} =-m$. 
We have the dynamical system
\begin{eqnarray}
\dot{m} &=& h +(\beta-1)m-\frac{m^3}{3} \\
\tau\dot{h} &=& -m
\end{eqnarray}
It is easy to see that for $\beta>1$ this system has no stable steady states.
Upon recurring to Lienard transformation  we have the equivalent second order system
\begin{equation}
\ddot{m} - (\beta-1-m^2)\dot{m} + \frac{1}{\tau} m = 0
\end{equation}
This is an instance of the Van der Pol equation \cite{van1934nonlinear}, that is known to display self oscillations  for $\beta>\beta_c=1$ and the phase transition has been substituted by an Andronov-Hopf bifurcation.  
This prediction has been  tested in the simplest microscopic setting, that is the fully connected Ising model. This is composed of  $N$ interacting spin variables $s_i=\pm 1$ whose energy  $E$ in an external magnetic field $h$ is 
($ M = \sum_i s_i $)
\begin{equation}
E = -J/2N   M^2 -hM
\end{equation}
and whose equilibrium configurations are given by the Boltzmann-Gibbs distribution
\begin{equation}
P({\bf s}) \propto \exp(-\beta E)
\end{equation}
We will consider a single spin flip heat bath dynamics, i.e. at each time step we choose one spin $s_i$ uniformly at random and set its new state, given the magnetization density $m=M/N$, with probability
\begin{eqnarray}
p(s_i=+1) = \frac{1}{e^{-\beta (J m+h)}+1} \\
p(s_i=-1) = \frac{1}{e^{\beta (J m+h)}+1}. 
\end{eqnarray}
and update the external field with the prescribed feedback rule ($\Delta t=1/N$)
\begin{equation}
 \Delta h = -m/(N\tau)
\end{equation}
Montecarlo simulations as well as approximate analytical calculations obtained by a Van Kampen system size expansion (see Appendix for further details) confirm very well the predictions, in particular 
\begin{itemize}
\item for $\beta\sim 1$, 
the system shows approximately harmonic oscillations  where the amplitude adjusts  itself making  the damping term negligible, i.e. $m^2 \sim \beta -1$.
\begin{figure}[h!!!!!]\label{fig2}
\begin{center}
\includegraphics*[width=0.75\textwidth,angle=0]{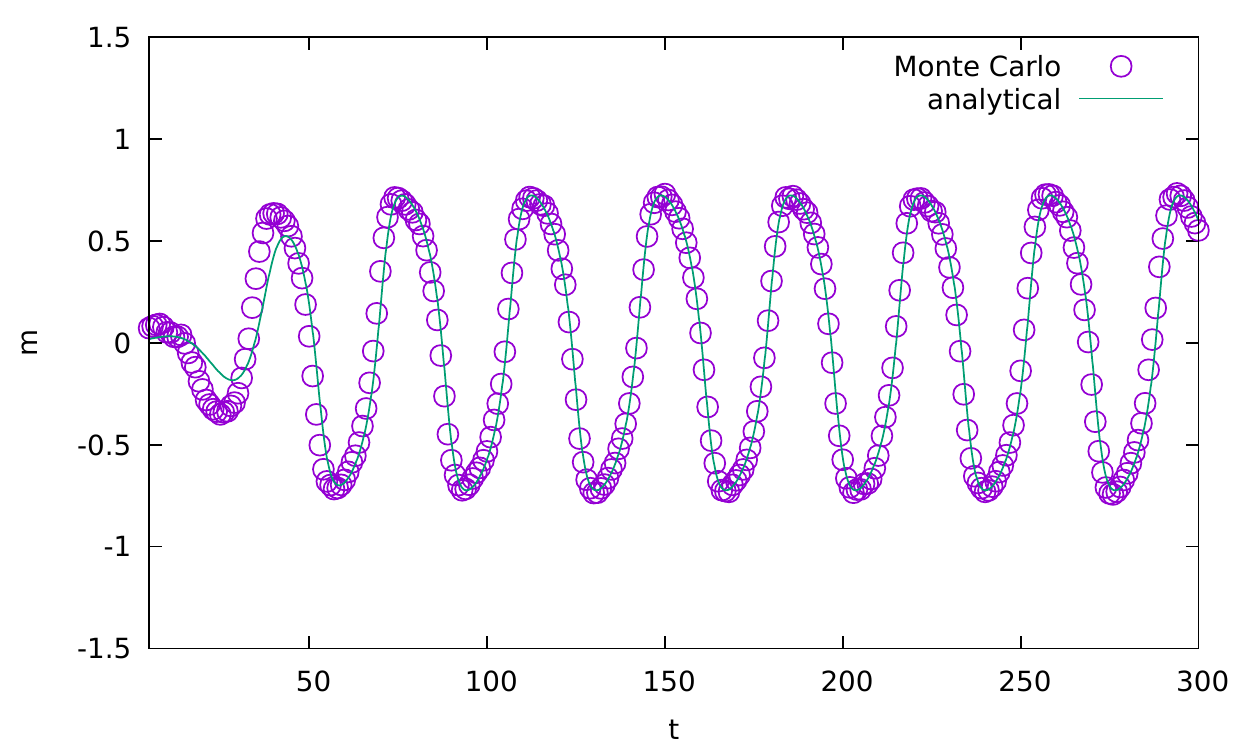}
\includegraphics*[width=0.75\textwidth,angle=0]{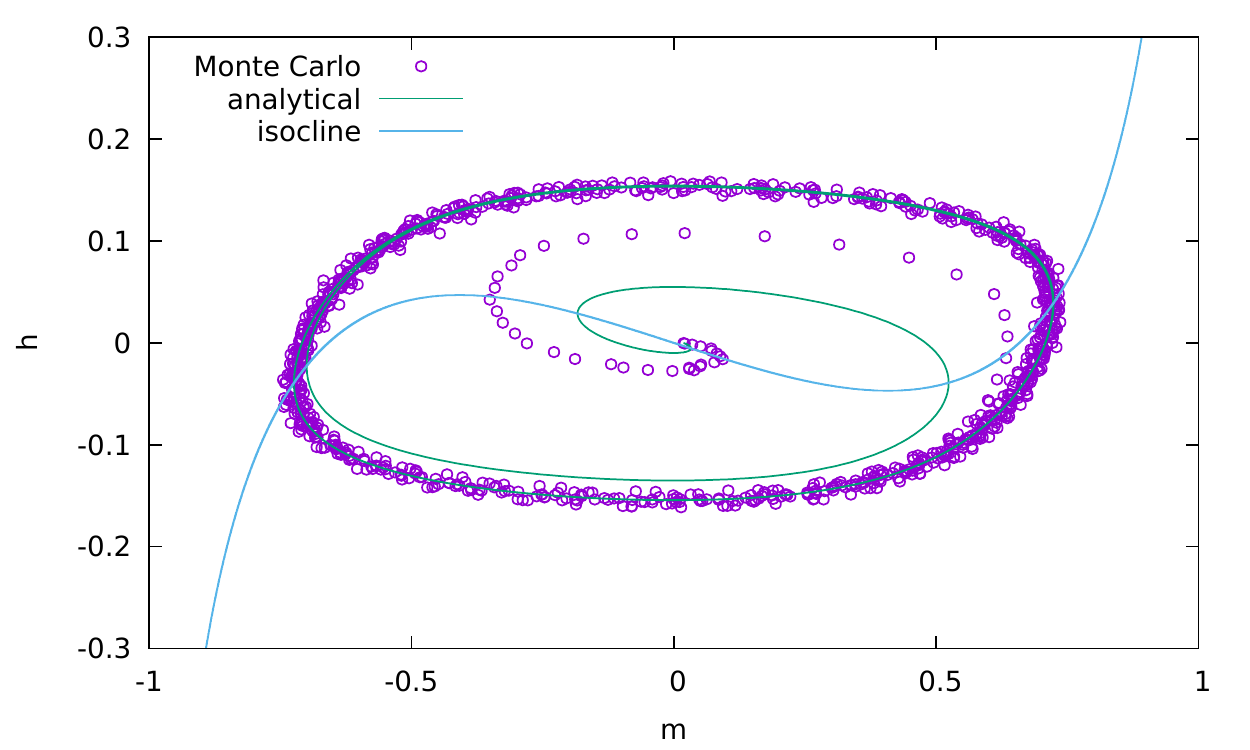}
\caption{Weakly sinusoidal oscillations in the fully connected Ising model, $\beta=1.2$, $\tau=30$, $N=10^4$. Top: Magnetization as a function of time. Bottom: Limit cycle in the plane $(m,h)$. From Monte Carlo simulations (points) and analytical calculations (lines).}
\end{center}
\end{figure}
\newpage
\item for $\beta\gg 1$ it performs relaxation oscillations: approximately, in the phase space $(m,h)$, dynamical trajectories follow the isocline curve  (where  $\frac{dh}{dm}\to\infty$) upto the tipping unstable points, where they follow straight horizontal lines corresponding to sudden jumps.

\begin{figure}[h!!!!!!!]\label{fig3}
\begin{center}
\includegraphics*[width=0.75\textwidth,angle=0]{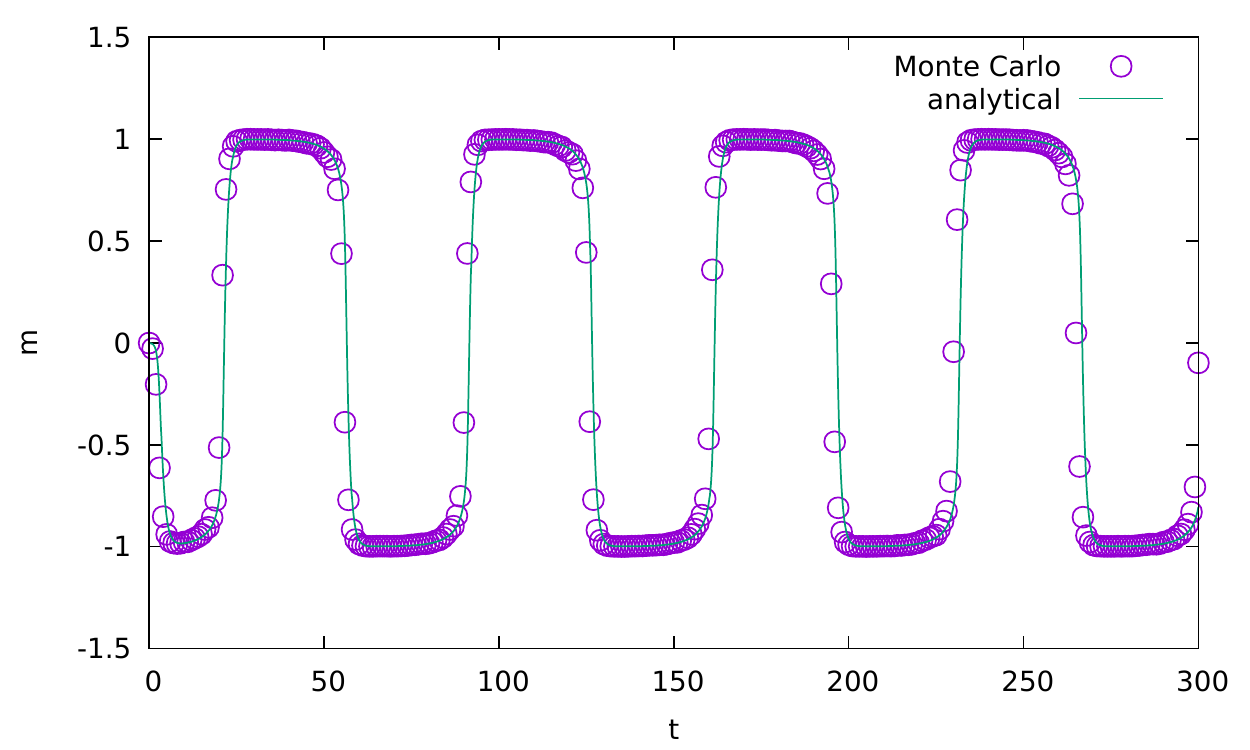}
\includegraphics*[width=0.75\textwidth,angle=0]{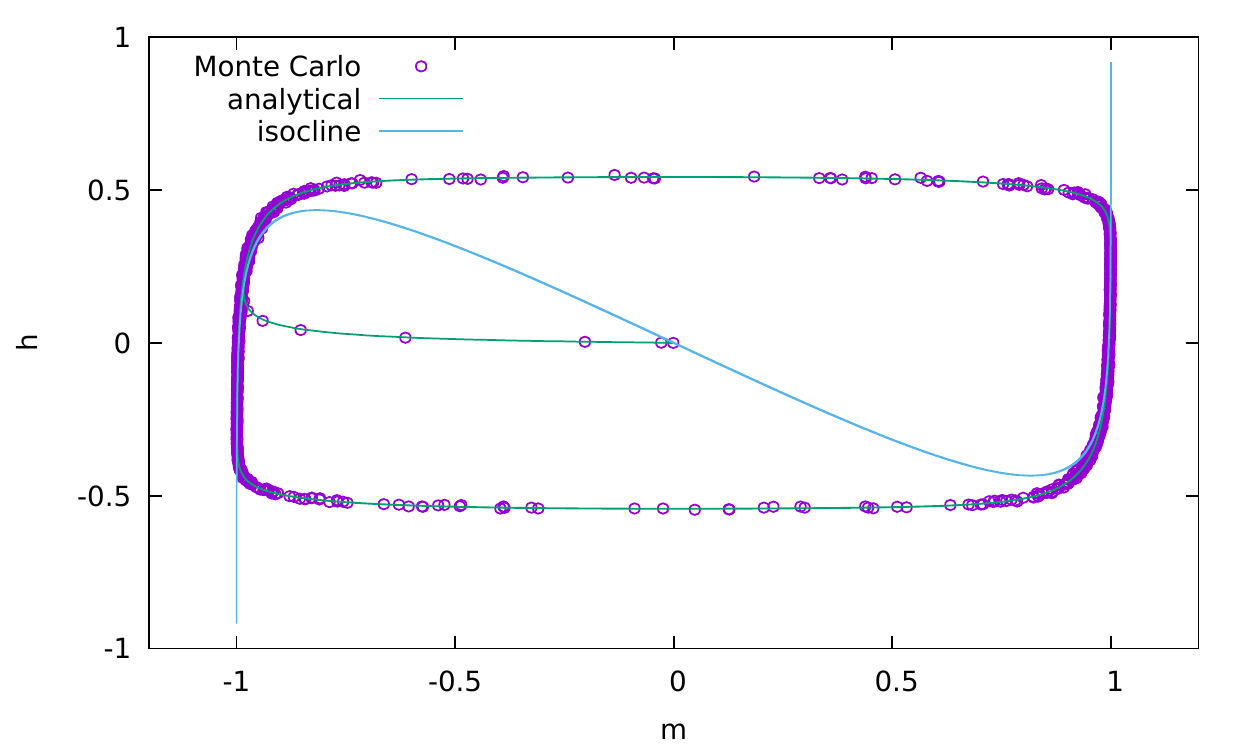}
\caption{Relaxation oscillations in the fully connected Ising model, $\beta=3$, $\tau=30$, $N=10^4$. Top: Magnetization as a function of time. Bottom: Limit cycle in the plane $(m,h)$.  From Monte Carlo simulations (points) and analytical calculations (lines)}
\end{center}
\end{figure}
 \end{itemize} 
Although the feedback has been introduced artificially here for illustrative purposes, the analyzed dynamics could be seen as an isothermal analogous of magnetic refrigeration cycles by magnetocaloric effect \cite{pecharsky1999magnetocaloric}.


\section*{Excess demand-price cycle: herding in a simple agent-based model}
The Ising model is a paradigm for the study of collective behaviors with applications extending beyond physics. In particular much research has been devoted in recent times in using concepts and methods from the statistical mechanics of the Ising model  to analyze agent-based models for social and economics phenomena \cite{de2006statistical}. In this section we will illustrate 
the emergence of business cycles \cite{geanakoplos2010leverage} in a simple market model of this kind as a consequence of the proposed theory.  
We will focus here  on a simple model  sharing similarity with minority games \cite{challet2013minority}, where $N$ agents shall choose one of two possible actions, i.e. the state of agent $i$  is given by a spin variable $s_i=\pm 1$.  
Upon identifying in a stylized way these two states as the propensity to sell or buy a given good, the magnetization $M=N_+-N_-$ is called the {\em excess demand}  and phenomenologically price  dynamics shall follow
(if more agents sell than buy, the price rises and viceversa) 
\begin{equation}
\dot{h} \propto -M
\end{equation}
where the price has been called $h$ in order to point out the analogies with  the Ising model and this equation is thus equivalent to the feedback  introduced  in the previous section. 
Given a reference price that we will set for simplicity to zero, we will assume that agents  update their state with a certain frequency $\nu$    
stochastically with probability (bounded rationality)
\begin{equation}
p(s=-1)/p(s=+1) = e^{-h}
\end{equation}
That takes into account the fact that agents have larger propensity to buy (sell) if they perceive that the price is below (above) the reference value. It can be easily seen that this simple model has a stable Gaussian fixed point. On the other hand, it is known that often economical and social agents can make decisions based  on imitation of neighbors in their social network. The cumulative effect of this imitation (herding) it is known to provide for instabilities in the underlying  dynamics \cite{sethna2001crackling}.
We will explore this issue of  network embedness and herding by recasting the aforementioned model in terms of the co-evolving network models studied in \cite{ehrhardt2006phenomenological}, where social links are created and destroyed among the agents, in turn reinforcing their beliefs, specifically:   
\begin{itemize}
\item Links are created with rate $\eta$ (per agent) among agents sharing the same state. 
\item Links are destroyed with rate $1$.
\end{itemize}
We will consider a regime of strong herding (that is the zero temperature limit considered as well in \cite{ehrhardt2006phenomenological}), i.e.
\begin{itemize}
\item Agents update their state upon looking at  the price only if they are isolated.
\end{itemize}
At fixed $h$ the equation of state of the system can be worked out approximately  with the methods of \cite{ehrhardt2006phenomenological}
(see the appendix for further details)
\begin{equation}
h = -\eta m +\log\frac{1+m}{1-m}
\end{equation}
that shows a second order critical point at $(\eta_c=2,h_c=0)$.
Upon considering phenomenologically the relaxation 
and considering the price feedback dynamics (with timescale $\tau$) we have finally the system
\begin{eqnarray}
\dot{m} &=& \eta m+h -\log\frac{1+m}{1-m} \\
\tau \dot{h} &=& -m
\end{eqnarray}
This prediction is tested againts Monte Carlo simulations where a good agreement is found (see fig. 4).
\begin{figure}[h!!!!!]\label{fig4}
\begin{center}
\includegraphics*[width=0.75\textwidth,angle=0]{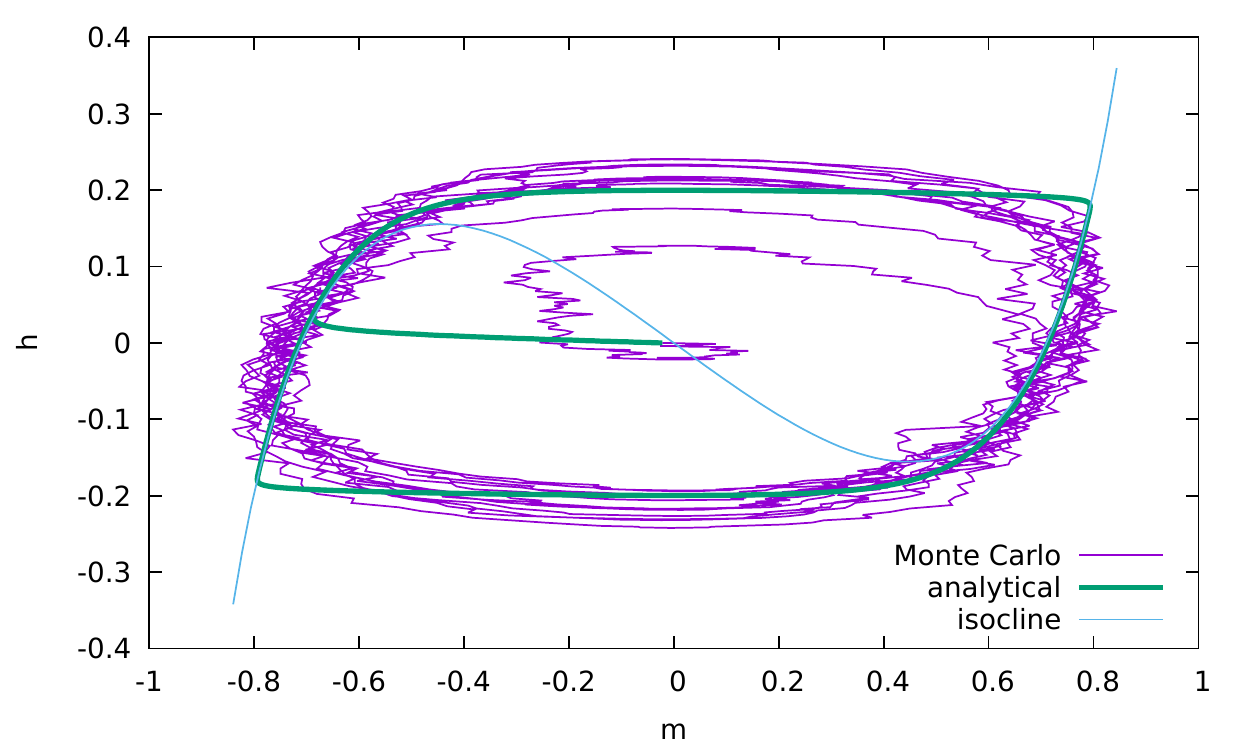}
\includegraphics*[width=0.75\textwidth,angle=0]{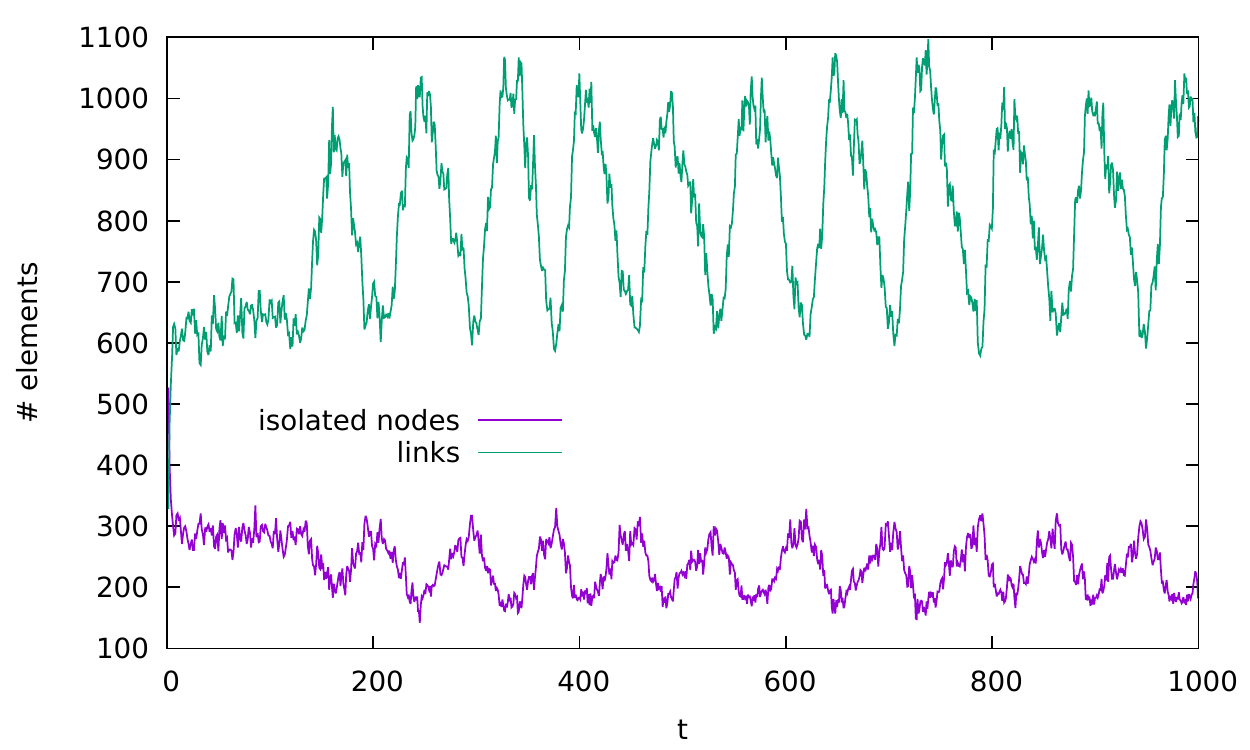}
\caption{Top: Self oscillations in the plane $(\textrm{excess demand } m, \textrm{price } h)$ for a simple agent-based market model in presence of herding. Bottom: social network oscillations, number of links  and isolated agents as a function of time. From Monte Carlo simulations and mean field analytics, $\eta=2.5$, $\tau=100$, $N=10^3$.}
\end{center}
\end{figure} 
Interestingly, in this case the whole network of interactions is oscillating in time, assuming the form of a Poissonian tree-like random graph (Erdos-Reny) whose average degree oscillates. This can be seen as an instance of a self-oscillating temporal network \cite{holme2012temporal}.

\section*{Congestion waves in queuing networks.}
In this section we will test the theory in the context of out-of-equilibrium systems. This is a very rich topic,  difficult to study given the lack of general established variational principles, like the maximum entropy, leading to the  Boltzmann-Gibbs measure characterizing their equilibrium counterpart. Perhaps one of the simplest yet general  class of processes displaying phase transitions in this context are the so-called zero range processes \cite{evans2005nonequilibrium}. Broadly speaking, these are models of particles hopping among nodes in a graph whose hopping rates depend only on the number of particles present on the departure and arrival nodes. This very short range interaction permits factorization of the steady state probability that is amenable for analytical calculations \cite{evans2005nonequilibrium}. Among them, we have   queuing networks \cite{kelly1976networks}, that are known to be subject to a congestion phase transition \cite{echenique2005dynamics, barankai2012effect}.
In these systems, used to model communication networks in engineering studies, packets of informations are injected into the network by users for processing purposes and stored in queues at the nodes. If the load that   the network experiences overcomes its processing capabilities, queues will start to grow congesting the system. In a stylized way, upon calling $n$ the number of packets in the network, $p$ the packet injection rate (``load'') and considering a function $f(n)$ that encodes for the network processing rate   we have the average rate equation:
\begin{equation}
\dot{n} = p - f(n)
\end{equation}
The function $f(n)$ depends on the chosen model, in general
$f(0)=0$, $f\geq 0$, $\lim_{n \to \infty}f(n)=\mu<\infty$. 
If $p> \mu$ we will have steady growing solution $\dot{n} \simeq p-\mu$, while 
stationary solutions are given by $n_s=f^{-1}(p)$ and are stable if $f'(n_s)>0$.
Generally unstable fixed points comes with coexisting stable ones in this setting. 

Traditional work on queuing network theory focused on the stationary regime while recent observations of traffic in large networks spurred an increasing interest study for the congested regime . However traffic data seem to show rather more complex phenomena with respect to simple steady growing queues, including waves and intermittent behaviors \cite{smith2011dynamics}, and it has been pointed out that one of the key ingredient accounting for the latters relies on the level of users feedback to overloaded conditions \cite{huberman1997social}. 
In the most simple setting this is naturally implemented  within our scheme by introducing a linear feedback dynamics for the load $p$:    users have typical reaction times $\tau$, they strive for  a desired load $p_d$ and they reduce the load if queues are longer (linear negative feedback of strength $c$):
\begin{eqnarray}
\dot{n} &=& p -f(n) \\
\tau \dot{p} &=& -(p-p_d) - c n
\end{eqnarray}
In presence of the feedback the dynamics could present no stable steady states but self-oscillations and it will be shown that this is the case in presence of a simple rule of traffic control.  Previous studies have shown that traffic control, mimicking known Internet protocols like the TCP/IP  can enhance the processing capabilities of the system, increasing the free flow region but at the price of introducing non-linearities that trigger the congestion transition in a discontinuous way, with hysteresis and coexistence \cite{de2009congestion, de2009minimal} that are reflected in a non-monotonous $f(n)$. In correspondence of these points we expect the linear feedback
to induce self oscillations, that shall be thus specific to cases in which traffic is controlled. 
\begin{figure}[h!!!!!]\label{fig5}
\begin{center}
\includegraphics*[width=0.75\textwidth,angle=0]{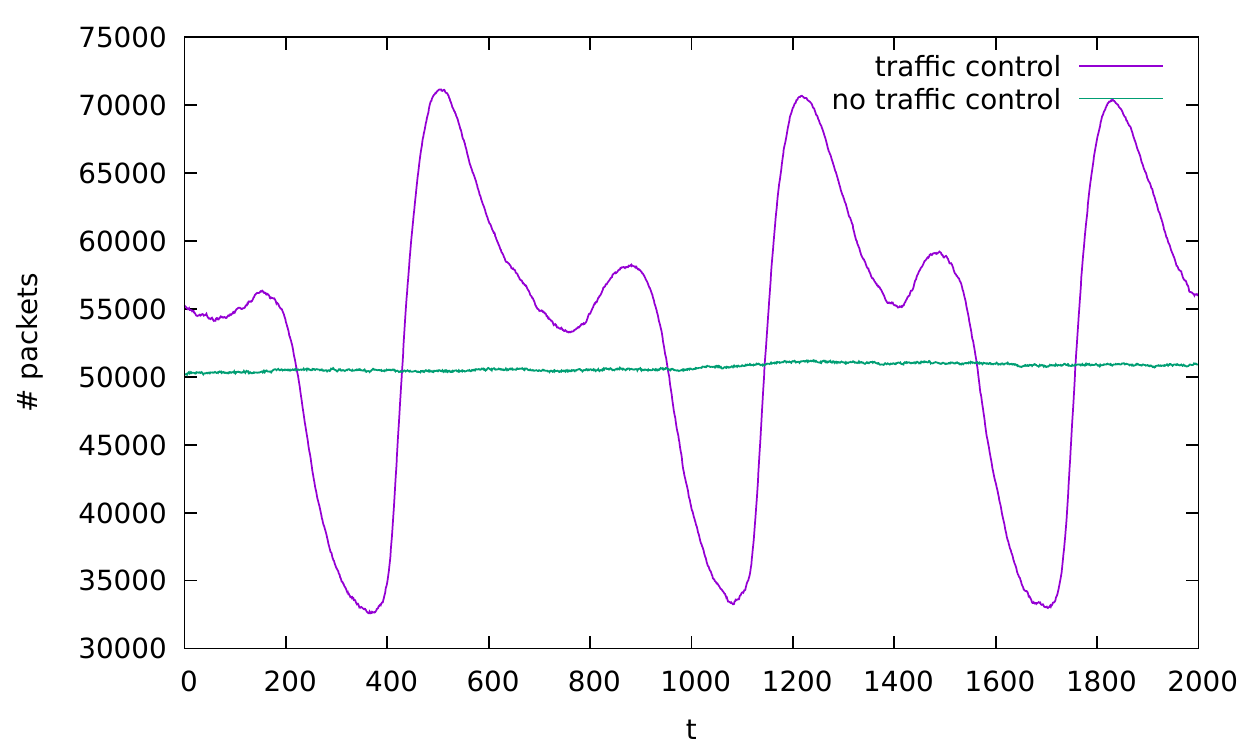}
\includegraphics*[width=0.75\textwidth,angle=0]{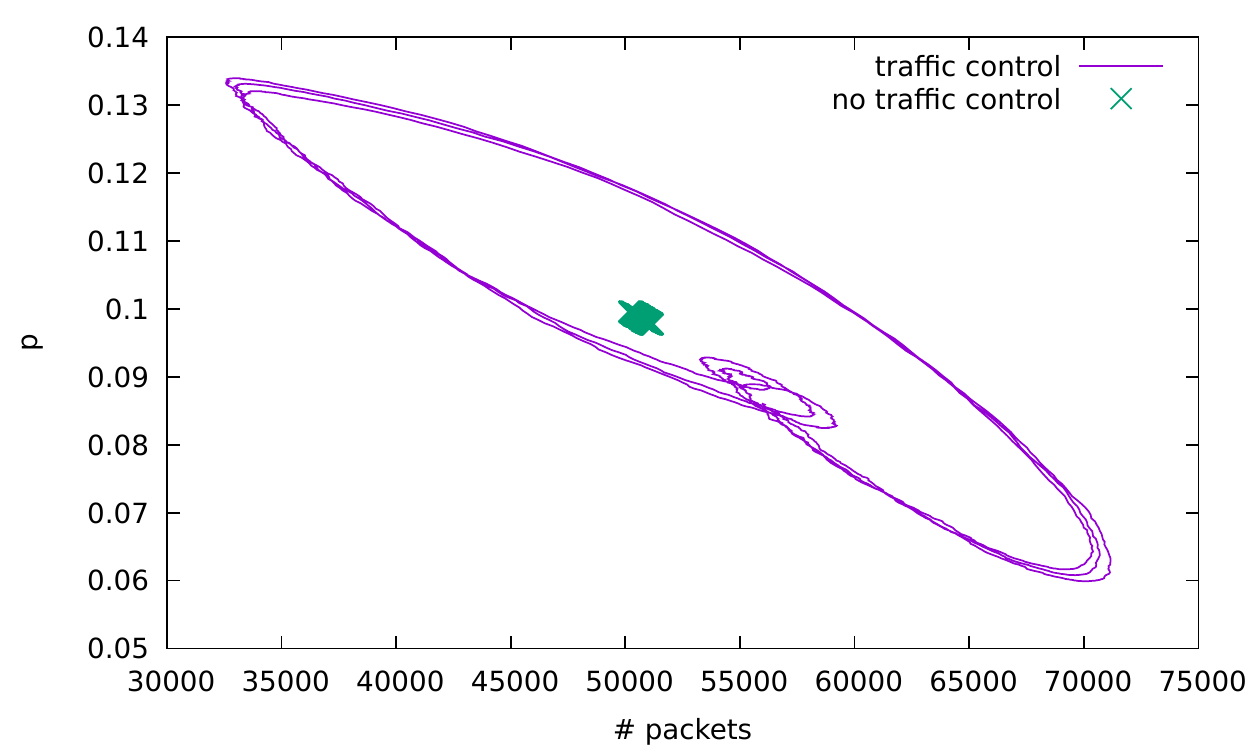}
\caption{Top: Number of packets as a function of time. Bottom: limit cycle in the plane $(n,p)$.  From Monte Carlo simulations on an Erdos-Renyi Jackson queuing network with and without traffic control, average degree $z=5$, $N=10^3$. In presence of traffic control the system develops self-oscillations.}
\end{center}
\end{figure}

These predictions have been succesfully tested on  large Jackson  queuing networks  (see  the appendix for further details).  
In presence of feedback dynamics for the $p$, we would expect now that upon loading the system by crossing  $f(n)$ in its decreasing part, this enforces self-oscillations and this is verified as we show in figure 5.


\section*{Autonomous metabolic network oscillations.}
In a biological context classical models in statistical physics are increasingly used to perform inference and analyze data: examples range from  random walks \& biopolymers \cite{de1979scaling}, Ising models \& neural networks \cite{humplik2017probabilistic} to  continuous spin models \& flocking birds \cite{cavagna2017physics} to cite few. Standard flux balance analysis (FBA) approaches 
to model cell metabolism \cite{orth2010flux} share a formal analogy with the Gardner problem in statistical mechanics \cite{gardner1988space} and  it has been recently shown that   maximum entropy  inference schemes \cite{de2016growth}  outperforms FBA in modeling flux data of the catabolic core of {\it E.coli} \cite{de2017statistical}. 
Metabolism is the network of enzymatic reactions that sustains the  free energy needs of the cell, strongly constrained by physico-chemical laws that in turn provide for suitable modeling.  Metabolic dynamics gives well-known  examples of non-linear self oscillators in particular glycolytic oscillations \cite{sel1968self}, experimentally tested  in living cells \cite{dano1999sustained}. 
What about whole cell metabolism? Recent findings in yeast show indeed    intrinsic whole single cell metabolic oscillations,  autonomous from the cell cycle and potentially able to  drive it \cite{papagiannakis2017autonomous}.  
In this section we will explore, within the proposed theory, the possibility that such oscillations are in general due to the effect of a feedback in presence of a bistable phenotypic landscape. 
Feedback  mechanisms are needed in order to mantain cell size homeostasis \cite{amir2014cell} and a bistable landscape has been observed in  {\it E. coli} \cite{deris2013innate, kotte2014phenotypic}, where it is at the core of  persistence phenomena \cite{balaban2004bacterial}.  Furthermore it has been recently pointed out theoretically within the framework of constraint-based models that second order moments constraints on the growth rate enable in general for bistability \cite{de2017maximum}. These are stationary models of  large chemical networks including in a realistic way the stoichiometry of known pathways and a phenomenological biomass growth reaction that is function of the enzymatic fluxes $\lambda = \lambda({\bf v})$. Within the feasible space  $P$  it is  possible to constrain the first two moments of the growth rate in the most unbiased way by recurring to the two parameters $(\beta,\gamma)$ Boltzmann distributions 
\begin{equation}
p({\bf v}) \propto \exp{(\beta \lambda({\bf v}) +\gamma \lambda({\bf v})^2)} \qquad {\bf v} \in P
\end{equation}
Upon counting the number of feasible states leading to the same growth rate by uniformly sampling $P$  the marginal growth rate distribution can be recasted in terms of the  rate function $F(\lambda)$ (where we posed $\lambda_{max}=1$, the maximum growth rate in the model obtainable by linear programming, and we get a simplex-like entropic term with $a\simeq 20$ for the carbon catabolic core of {\it E.Coli}) 
\begin{eqnarray}
p(\lambda) \propto e^{F(\lambda)} \\
F(\lambda) = \beta \lambda +\gamma \lambda^2+a\log(1-\lambda)  
\end{eqnarray}
It shall be noticed that such maximum entropy distribution can be the steady state of a suited population dynamics, that for the case $\gamma=0$ has been shown to be the logistic \cite{PhysRevE.96.010401}. If we consider relaxation dynamics in the linear reponse regime and linear control through $\beta$, looking for a desired $\lambda_s$ , we have the dynamical system
\begin{eqnarray}
\dot{\lambda} = \frac{\partial F}{\partial \lambda} = \beta +2 \gamma \lambda -\frac{a}{1-\lambda} \\
\tau \dot{\beta} = -(\lambda-\lambda_s) 
\end{eqnarray}
that upon Lienard trasformation is mapped into the second order system
\begin{equation}
\ddot{\lambda} -(2 \gamma -\frac{a}{(1-\lambda^2)})\dot{\lambda} +\frac{\lambda-\lambda_s}{\tau}=0
\end{equation}
A sufficient condition to get sel-oscillations  is
 $\gamma>\frac{a}{2(1-\lambda_s)^2}$. 
\begin{figure}[h!!!!!]\label{fig6}
\begin{center}
\includegraphics*[width=0.75\textwidth,angle=0]{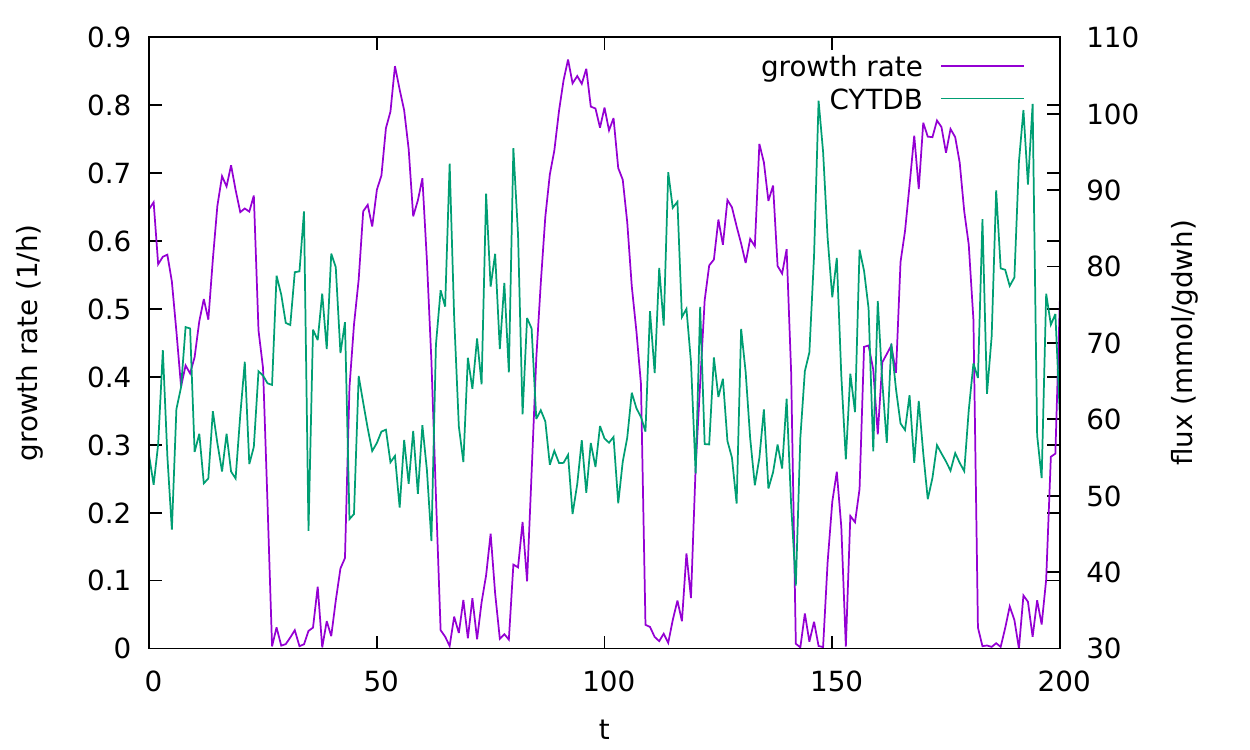}
\includegraphics*[width=0.7\textwidth,angle=0]{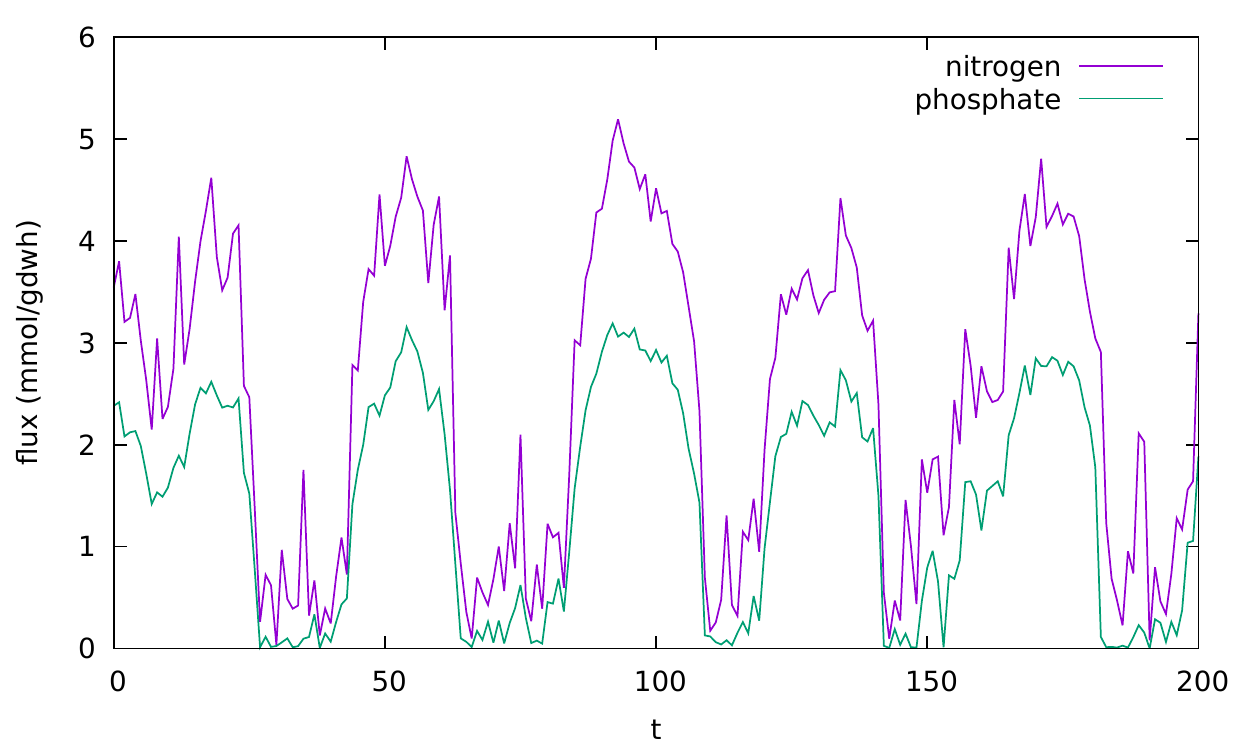}
\caption{Top: growth rate and cytochrome b activity as a function of time. Bottom:  uptakes (nitrogen and phosphate) as a function of time. From Montecarlo simulations of a maximum entropy model of the carbon catabolic core of {\it E. coli} in presence of feedback for growth control in a bistable phenotypic landscape ($\gamma=50$, $\tau=100$, $\lambda_s=0.4$h$^{-1}$). }
\end{center}
\end{figure} 
This is confirmed by simulations on a model of the catabolic core of {\it E coli} (Fig 6, see appendix for further details), where growth rate oscillations entrain correlated metabolic fluxes. Here are shown  the activity of the enzyme Cytochrome b oxidase and the nitrogen and phosphate uptakes.

\section*{Conclusions}
In this work we gave a general prescription for the emergence of self-oscillations in large systems with many interacting units:  a negative linear feedback between the control and order parameters in presence of phase coexistence. The oscillatory collective  behavior emergent in this way does not depend on postulating oscillatory units, but it is fully self-organized.
The key idea is that large systems endowing phase coexistence develop self-oscillations in presence of feedback that try to force them  on thermodynamically unstable branches, analogously to  active electrical devices controlled with a workload corresponding to ``negative resistance'' parts of their characteristic curve. 
It has been shown that  the feedback maps the Landau mean-field theory  into the Van-der-Pol oscillator and such behavior has been confirmed for the fully connected Ising model subject to an heat bath dynamics. Further studies are needed to test the theory on finite dimensions as well as experimentally. Preliminary numerical results on a $2d$ square lattice seem to show that self-oscillations are triggered by local feedback, while a global one triggers  separation of magnetic domains.  The theory is  general and here it has been applied to describe in a stylized way i) excess demand-price cycles due to strong herding effects in  a simple agent-based market model; ii) congestion waves in  queuing networks  triggered by users feedback to delays in overloaded conditions; iii)  metabolic network oscillations resulting from cell growth control in a bistable phenotypic landscape.   All of them would deserve further work by their own, in particular suited  analysis to test them against data by means of promising ICA-based methods \cite{de2005complexity, ciaramella2006ica}.  
The theory could open the way to explore the  thermodynamics \& statistical physics of self oscillations,  and recent tools developed within stochastic thermodynamics could play a key role in this respect \cite{seifert2012stochastic, zhang2016critical}. Finally, within the framework of control theory self-oscillations can be seen as an unwanted negative side effect  and this theory could help shedding light  on the origin of this problem while controlling large complex networks \cite{liu2011controllability}.

\section*{Appendix}
\subsection*{Van Kampen expansion for the fully connected Ising model}
Upon considering the induced evolution for the magnetization $M$ from the heat bath dynamics of the single spins, this performs a random walk in the interval $[-N,N]$ with stepsize $-2,0,2$ and  the (normalized) rates
\begin{itemize}
\item $   W(M \to M+2)=  N \frac{1-m}{2}\frac{1}{e^{-\beta (J m+h)}+1} \theta(1-m)$
\item $    W(M \to M-2)=  N \frac{1+m}{2}\frac{1}{e^{\beta (J m+h)}+1} \theta(1+m)$
\item $W(M \to M) = N-W(M \to M-2)-W(M \to M+2)$
\end{itemize}
The master equation
\begin{eqnarray}
\dot{P}(M)  = -P(M) ( W(M \to M-2) +  W(M \to M+2) ) \nonumber \\
+ P(M-2) W(M-2 \to M) + P(M+2) W (M+2 \to M)
\end{eqnarray}
can be expanded in the system size \cite{kampen1961power}, i.e. upon considering a decomposition of $M$ under a scaling hypothesis in term of the auxilary variables
\begin{equation}
M = N \phi +\sqrt{N} z,
\end{equation}
performing an expansion of the master equation in the parameter $1/\sqrt{N}$ and neglecting higher order terms, one have, upon considering
\begin{eqnarray}
W_+ = \frac{1-\phi}{1+e^{-\beta (J \phi+h)}} \quad W_- = \frac{1+\phi}{1+e^{\beta (J \phi+h)}} \\
\alpha_1(\phi) =  W_+ - W_- \quad \alpha_2(\phi) = 2 (W_+ + W_-) 
\end{eqnarray} 
on one hand a deterministic equation for the $\phi$
\begin{equation}
\dot{\phi} = \alpha_1(\phi)
\end{equation}
on the other a linear Fokker-Planck equation for the $z$
\begin{equation}
\frac{\partial P(z)}{\partial t} = -\alpha'_1(\phi) \frac{ \partial (z P(z))}{\partial z} + \frac{1}{2}\alpha_2 (\phi) \frac{\partial^2 P(z)}{\partial z^2}
\end{equation}
where both depend on the external parameter $h$, that can be considered varying in time and subject to the negative feedback
\begin{equation}
\tau \dot{h} = -\phi
\end{equation}

\subsection*{Mean field derivation of the equation of state of the herding model.}
We will recur to the mean field techniques defined in \cite{ehrhardt2006phenomenological}. 
The approximate population dynamics equations for the density of agents with $k$ connections in state $\sigma$ read:
\begin{eqnarray}
\dot{n}_{k,\sigma} &=& (k+1)n_{k+1,\sigma}-k n_{k,\sigma} +x_\sigma (n_{k-1,\sigma}- n_{k,\sigma}) \quad k>0 \\
x_\sigma &=& \eta\sum_k n_{k,\sigma} \\
\dot{n}_{0,\sigma} &=& n_{1,\sigma}-x_\sigma n_{0,\sigma} +\nu_{-\sigma} n_{0,-\sigma}-\nu_{\sigma}n_{0,\sigma}
\end{eqnarray}
If we consider the generating function $G_\sigma(s)=\sum_k n_{k,\sigma}s^k$ we have in the steady state $G_\sigma^{ss}(s)=n_{0,\sigma}e^{x_\sigma s}$ and the self consistent equations
\begin{eqnarray}
\nu_{+}n_{0,+}=\nu_{-}n_{0,-} \\
x_{\sigma}/\eta= n_{0,\sigma}e^{x_\sigma} \\
x_++x_-=\eta
\end{eqnarray}
If we parametrize $\nu_-/\nu_+=e^h$ and consider $m=\frac{x_+-x_-}{\eta}$
we have finally the equation of state (where $\eta_c=2$)
\begin{equation}
h = -\eta m +\log\frac{1+m}{1-m}
\end{equation}

\subsection*{Queuing network model}
The  model employed here is the  Jackson or open queuing network, consisting  of $N$ nodes such that:
\begin{itemize}
\item each node $i$  is endowed with a FIFO (first-in first out) queue
      with unlimited waiting places (it can be arbitrary long).
\item The delivery of a packet from the front of $i$ follows a poisson process with a certain frequency $t_i$(service rates), and
      \item[-] the packet exits the network with some probability $\mu_i$, or
      \item[-] it goes on the ``back'' of another queue $j$ with probability $q_{ij}$.
\item Packets are injected in each queue $i$ from external sources
      by a Poisson stream with intensity $p_i$.
\end{itemize}
We considered random walk routing  $q_{ij}=1/k_j$ where $k_j$ is the degree of the receiving node $j$, and completely homogeneous conditions $p_i=p$, $\mu_i=\mu$, $t_i=t=1$.
Traffic control has been included with the following simple rule \cite{de2009congestion}:
\begin{itemize}
\item The receiving node $j$ starts to reject particles with probability  $\eta$ once its queue is longer than $n^*$
\end{itemize} 
Results shown in Figure 5 are obtained by simulations on an Erdos-Reny random graph (average degree $z=5$, size $N=10^3$),  desired load  \& absorbing rate $p_d=\mu=0.2$, user feedback strength $c=2\cdot 10^{-6}$, and traffic control parameters $\eta=0.75$, $n^*=10$.

\subsection*{Maximum entropy constraint based models of metabolic networks}
In constraint-based modeling a metabolic  system is modeled in terms of the dynamics of the concentration levels and reaction fluxes, under the assumption of well-mixing, steady state and neglecting molecular noise.
For a chemical reaction  network  in which $M$ metabolites participate in $N$ reactions  with the  stoichiometry encoded in a matrix $\mathbf{S}=\{S_{\mu r}\}$, the concentrations $c_\mu$ change in time according to mass-balance equations
\begin{equation}
\dot{\mathbf{c}} = \mathbf{S \cdot v}
\end{equation}
where $v_i$ is the flux of the reaction $i$ (that is in general a  function of the concentration levels  $v_i(\mathbf{c})$). The steady state implies
$\mathbf{S \cdot v}=0$. In constraints-based modeling, apart from mass balance constraints, fluxes are bounded in certain ranges $v_r \in [v_{r}^{{\rm min}},v_{r}^{{\rm max}}]$ that take into account thermodynamic irreversibility, kinetic limits and physiological constraints.
The set of constraints
\begin{eqnarray}\label{eq3}
\mathbf{S \cdot v}=0, \nonumber \\
v_r \in [v_{r}^{{\rm min}},v_{r}^{{\rm max}}]
\end{eqnarray}  
defines  a convex polytope $P$ in the space of reaction fluxes. We seek for the states fixing the first two moments of the growth rate, given by the Boltzmann distributions:
\begin{equation}
p({\bf v}) \propto \exp{(\beta \lambda({\bf v}) +\gamma \lambda({\bf v})^2)} \qquad {\bf v} \in P
\end{equation}
They have been sampled  by means of an hit-and-run Monte Carlo Markov chain with ellipsoidal rounding \cite{de2015uniform} on the model of the catabolic core of {\it E. coli} from the genome scale reconstruction \cite{orth2011comprehensive},  including glycolysis, pentose phosphate pathway, Krebbs cycle, oxidative phosphorylation, nitrogen catabolism and the biomass growth reaction, for a total of $N=95$ reactions and $M=72$ compounds, simulated in a glucose limited aerobic minimal medium with a maximum glucose uptake of $u_{g,max}=10$mmol/gdwh,

\section*{Acknowledgments}
The author thanks S.De Martino, M.Falanga, G. Tkacik \& M. Lang for interesting discussions. 
The research leading to these results has received funding from the People Programme (Marie Curie Actions) of the European Union's Seventh Framework Programme ($FP7/2007-2013$) under REA grant agreement $n[291734]$.

\bibliographystyle{ieeetr}

\bibliography{ref_osc}

\end{document}